%% file: 0Main_Paper.tex
\documentclass[letterpaper, 10 pt, conference]{ieeeconf}

\input{0Config.tex}
\input{0Glossary.tex}

\allowdisplaybreaks

\title{\LARGE \bf Model-free Reinforcement Learning for \\Non-stationary Mean Field Games}
\author{Rajesh K Mishra, Deepanshu Vasal, and Sriram Vishwanath}

\begin{document}

	\maketitle

	\begin{abstract}
		In this paper, we consider a finite horizon, non-stationary, \mfg with a large population of homogeneous players, sequentially making strategic decisions, where each player is affected by other players through an aggregate population state termed as \emph{mean field state}. Each player has a private type that only it can observe, and a mean field population state representing the empirical distribution of other players' types, which is shared among all of them. Recently, authors in~\cite{Vasal2019} provided a sequential decomposition algorithm to compute \mfe for such games which allows for the computation of equilibrium policies for them in linear time than exponential, as before. In this paper, we extend it for the case when state transitions are not known, to propose a reinforcement learning algorithm based on Expected Sarsa with a policy gradient approach that learns the \mfe policy by learning the dynamics of the game simultaneously. We illustrate our results using cyber-physical security example.
	
	\end{abstract}

	\section{Introduction}
		\input{1Introduction.tex}
	\section{Model}
		\label{sec:Model}
		\input{2Model.tex}

	\section{Preliminaries}
		\label{sec:Prelims}
		\input{3Preliminaries.tex}
	\section{Reinforcement Algorithm}
		\label{sec:Algo}
		\input{4Algorithm.tex}

	\section{Convergence}
		\input{5Convergence.tex}

	\section{Numerical Example}
		\label{sec:Prblm}
		\input{6Numerical_problems.tex}

	\section{Conclusion}
		\label{sec:Concl}

\input{7Conclusion.tex}

	\medskip
	\small
	\bibliographystyle{IEEEtran}
	\bibliography{References}
\end{document}

%% file: 0Config.tex
\usepackage{bbm}
\usepackage{amsmath,amssymb,amsfonts}
\usepackage{mathrsfs}
\usepackage{xcolor}
\usepackage{graphicx}
\usepackage{amsfonts}
\usepackage[english]{babel}
\usepackage{epstopdf}
\usepackage{textcomp}
\usepackage{mathtools}
\usepackage[linesnumbered,ruled,vlined]{algorithm2e}
\usepackage{dsfont}
\newtheorem{theorem}{Theorem}

\newtheorem{lemma}{Lemma}

\newtheorem{lemma*}{Lemma}
\newtheorem{assumption}{Assumption}

\renewcommand{\hat}{\widehat}

\newcommand{\cN}{\mathcal{N}}
\newcommand{\cT}{\mathcal{T}}

\newcommand{\nn}{\nonumber}    

\newcommand{\defeq}{\buildrel\triangle\over =}
\newcommand{\sq}[1]{\left[#1\right]}
\newcommand{\cm}[1]{\left(#1\right)}
\newcommand{\bm}[1]{\left\{#1\right\}}

\newcommand{\mE}{\mathbb{E}}
\newcommand{\mG}{\mathbb{G}}
\newcommand{\cX}{\mathcal{X}}
\newcommand{\cZ}{\mathcal{Z}}

\newcommand{\cA}{\mathcal{A}}

\newcommand{\cK}{\mathcal{K}}

\newcommand{\cP}{\mathcal{P}}

\newcommand{\cH}{\mathcal{H}}
\newcommand{\cS}{\mathcal{S}}

\newcommand{\tgamma}{\tilde{\gamma}}
\newcommand{\tsigma}{\tilde{\sigma}}

\newcommand{\ttheta}{\tilde{\theta}}

\newcommand{\figref}[1]{Figure~\ref{#1}}

%% file: 0Glossary.tex
\usepackage{xspace}
\usepackage[acronym,nogroupskip,nonumberlist,nopostdot]{glossaries}
\glsdisablehyper

\newacronym{mdp}{MDP}{Markov decision process}
\newacronym{ne}{NE}{Nash equillibrium}
\newacronym{mfe}{MFE}{mean field equillibrium}
\newacronym{mpe}{MPE}{Markov perfect equillibrium}
\newacronym{mfg}{MFG}{mean field game}
\newacronym{rl}{RL}{reinforcement learning}
\newacronym{marl}{MARL}{multi-agent reinforcement learning}
\newacronym{iot}{IoT}{Internet of Things}
\newacronym{ssg}{SSG}{Stackelberg security game}
\newacronym{pse}{PSE}{perfect Stackelberg equilibrium}
\newacronym{mpse}{MPSE}{Markov \gls{pse}}
\newacronym{pomdp}{POMDP}{partially observed \gls{mdp}}
\newacronym{mc}{MC}{Monte Carlo}
\newacronym{pbe}{PBE}{perfect Bayesian equilibrium}
\newacronym{gmfg}{GMFG}{graphon mean field game}
\newacronym{gmfe}{GMFE}{graphon mean field equillibrium}
\newacronym{spe}{SPE}{sub-game perfect equillibrium}
\newacronym{mkv}{MKV}{McKean-Vlasov}

\newcommand{\mdp}{\gls{mdp}\xspace}
\newcommand{\gne}{\gls{ne}\xspace}
\newcommand{\mfe}{\gls{mfe}\xspace}
\newcommand{\mpe}{\gls{mpe}\xspace}
\newcommand{\mfg}{\gls{mfg}\xspace}
\newcommand{\mfgs}{\glspl{mfg}\xspace}
\newcommand{\rl}{\gls{rl}\xspace}
\newcommand{\marl}{\gls{marl}\xspace}

%% file: 1Introduction.tex
There is an increasing number of applications today that involve \emph{large scale} interactions among strategic agents, such as smart grid, autonomous vehicles, cyber-physical systems, \gls{iot}, renewable energy markets, electric vehicle charging, ride sharing apps, financial markets, crypto-currencies and many more. Such applications can be modeled through \mfgs introduced by Huang et al~\cite{Huang2006} and Lasry and Lions~\cite{Lasry2007}, where each player is affected by other players not individually, but through a `mean field' that is an aggregate of the population states. 

Multi-agent systems have become ubiquitous owing to their variety of applications. In the recent decade, several \marl approaches have been proposed to learn optimal strategies for the agents in the system. However, these approaches scale poorly with size and the dynamic nature of the environment makes the learning of these strategies rather difficult. In contrast, a planning framework based on the mean field approximation has shown potential due to the remarkable property wherein the agents decouple from one another with the increase in their number. As such the agents interact only through the mean field which makes dynamics of the system tractable. \marl systems work well with probabilistic models but fail with large number of agents and this is where the mean field concept has proven useful. 

mean field approximation has been in use in game theory applications to study large number of non-cooperative players. In such games, given the mean field states, players' \mfe strategies are computed backward recursively through dynamic programming (or HJB equation in continuous time), whereas given players' strategies, mean field states are computed forward recursively using Mckean Vlasov equation (or Fokker-Plank equation). Overall, \mfe strategies and mean field states are coupled across time through a fixed-point equation. Recently, in~\cite{Vasal2019}, the author presented a sequential decomposition algorithm that computes such equilibrium strategies by decomposing this fixed-point equation across time, and reducing the complexity of this problem from exponential to linear in time. It involves solving a smaller fixed-point equation for each time instant $t$.
 
The concept of a generalized \mfg is discussed by authors in~\cite{guo2019learning}, where they prove the uniqueness and existence of \gne. They also propose a Q-learning algorithm with Boltzmann policy and analyze its convergence properties and complexity. In~\cite{tiwari2019reinforcement}, the authors propose a posterior sampling based approach where each agent samples a transition probability from a previous transition and converges to the optimal oblivious strategy for \mfgs. Authors in~\cite{yang2018mean} consider a multi-agent setting with agents coupled by the average action of the agents. They show that the mean field Q algorithm that converges to the \gne. In~\cite{Subramanian2019a}, the authors propose \rl algorithms for stationary \mfgs that compute \mfe and social-welfare optimal solution strategies. In~\cite{elie2019approximate}, the authors consider a non stationary \mfg and propose a fictitious play iterative learning algorithm to devise optimal strategies for mean field states. They argue that if the agent plays the best response to the observed population state flow, they eventually converge to the \gne.

Most prior research assume a full knowledge of the dynamics of the system. They also assume a stationary mean field approximation while computing the \mfe. In our paper, however, we consider a non stationary \mfg with no prior knowledge of the system dynamics, to derive the optimal policy as a function of the mean field state. The main contribution of our paper is an \rl algorithm that solves the fixed point equation that was proposed by the authors in~\cite{Vasal2019} while learning the dynamics of the \mdp simultaneously. We  prove the convergence of our algorithm to the \mfe analytically. Finally, we show convergence to the mean field equilibrium strategy for a stylized problem of \emph{Malware Spread} where the strategy derived from our algorithm coincides with the optimal strategy obtained assuming the full knowledge of the system. 

The paper is structured as follows. In Section~\ref{sec:Model}, we present our model. In Section~\ref{sec:Prelims}, we present the sequential decomposition algorithm with backward recursion presented in~\cite{Vasal2019} to compute \mfe of the game. In Section~\ref{sec:Algo}, we present our reinforcement learning algorithm and prove its convergence to \mfe polices. In Section~\ref{sec:Prblm}, we present a cyber-physical system security example. We conclude in Section~\ref{sec:Concl}.

\subsection{Notation}

We use uppercase letters for random variables and lowercase for their realizations. For any variable, subscripts represent time indices and superscripts represent player identities. We use notation $-i$ to denote all players other than the player $i$ i.e. $-i=\left\{1,2,\ldots,i-1,i+1,\ldots,N\right\}$. We use notation $a_{t:t\prime}$ to represent the vector $\left(a_t,a_t+1,\ldots,a_{t^\prime}\right)$ when $t^\prime \geq t$ or an empty vector if $t^\prime <t$. We use $a_t^{-i}$ to mean $\left(a_t^1, a_t^2,\ldots, a_t^{i-1}, a_t^{i+1}, \ldots, a_t^N\right)$. We remove superscripts or subscripts if we want to represent the whole vector, for example $a_t$  represents $(a_t^1, \ldots, a_t^N) $. We denote the indicator function of any set $A$ by $\mathds{1}_{\{A\}}$. For any finite set $\mathcal{S}$, $\mathcal{P}(\mathcal{S})$ represents space of probability measures on $\mathcal{S}$ and $|\mathcal{S}|$ represents its cardinality. We denote by $P^{\sigma}$ (or $E^{\sigma}$) the probability measure generated by (or expectation with respect to) strategy profile $\sigma$ and the space for all such strategies as $\mathcal{K}^\sigma$. We denote the set of real numbers by $\mathbb{R}$. All equalities and inequalities involving random variables are to be interpreted in \emph{a.s.} sense. 

%% file: 2Model.tex
We consider a finite horizon discrete-time large population sequential game with $N$ homogeneous players, where $N$ tends to $\infty$. We denote the set of homogeneous players by $\cN$ and the set of time periods by $\cT$. In each time instant $t\in\cT$, a player $i\in\cN$ observes a private type $x_t^i\in\cX = \left\{1,2,\cdots, N_x\right\}$ and a common observation of the mean field population state $z_t\in\cZ$, where $z_t = \left(z_t(1),z_t(2),\ldots,z_t(N_x)\right)$ is the fraction of population having type $x\in\cX$ at time $t$ given as.
\begin{align}
	z_t(x) = \frac{1}{N}\sum_{i=1}^N \mathds{1}_{\bm{x_t^i = x}}
\end{align} 
with $\sum_{i=1}^{N_x} z_t(i) = 1$. Consequently, the player $i$ takes an action $a_t^i\in\cA = \left\{1,2,\cdots, N_a\right\}$ based on policy $\sigma_t^i\left(.|z_{1:t},x_t^i\right)$, and receives a reward $R\left(x_t^i,a_t^i,z_t\right)$ which is a function of its current type $x_t^i$, action $a_t^i$ and the common observation $z_t$. Player $i$'s type evolves as a controlled Markov process,
\begin{align}
	x_{t+1}^i = f_x\left(x_t^i, a_t^i, z_t, w_t^i\right).
\end{align}
The random variables $(w_t^i)_{i\in\cN,t\in\cT}$ are assumed to be mutually independent across players and across time. We can also express the above update of $x_t^i$ through a kernel, $x_{t+1}^i\sim \tau(\cdot|x_t^i, a_t^i, z_t)$, where $\tau\left(\cdot\vert\cdot\right)$ represents the transition probabilities of the \mdp. In this paper, we assume that $\tau$ is unknown. We develop a model-free \rl algorithm to derive the equilibrium policy where $\tau$ is used to simulate the model and generate samples for learning. The idea of \mfgs is to approximate the finite population game with an infinite population game such the the mean field state $z_t$ converges to the statistical \mfe of the game~\cite{Subramanian2019a}.
 
In any time period $t$, player $i$ observes $(z_{1:t},x_{1:t}^i)$ and takes action $a_t^i$ according to a behavioral strategy $\sigma^i = (\sigma_t^i)_t$, where $\sigma_t^i:\cZ^{t}\times\mathcal{X}^t \to \mathcal{P}(\mathcal{A})$ defined over the space $\mathcal{K}^{\sigma}$ which implies $A_t^i\sim \sigma_t^i(\cdot|z_{1:t},x_{1:t}^i)$. In the case of a finite time-horizon game, $\mathbb{G}_T$, each player wants to choose the strategy $\tsigma$ that maximizes its total expected discounted reward over a time-horizon $T$, discounted by discount factor $0<\delta\leq 1$, given as
\begin{align}
	\label{Eqn:j_func}
	J^{\tsigma}_t=\mathbb{E}^{\tsigma_{t:T}}\left[\sum_{k=t}^{T}\delta^{k-t}R\left(X_k^i,A_k^i,z_k\right)\vert z_t,x_t^i\right].
\end{align}

%% file: 3Preliminaries.tex
\subsection{Solution concept: \mpe}
	The Nash equilibrium of $\mG_{T}$ is defined as strategies $\tsigma = (\tsigma_t^i)_{i\in\cN,t\in\cT}$ that satisfy, for all $i\in\cN$, 
	\begin{align}	
		\label{Eqn:mpe}
			\mE^{\tsigma^i,\tsigma^{-i}}&\left[\sum_{t=1}^T \delta^{t-1} R\left(X_t^i,A_t^i,Z_t\right) \right] \nn 
			\\ &\geq \mE^{\sigma^i,\tsigma^{-i}}\left[\sum_{t=1}^T \delta^{t-1} R\left(X_t^i,A_t^i,Z_t\right)\right].	
		\end{align}

	For sequential games, however, a more appropriate equilibrium concept is \mpe~\cite{Maskin2001}, which is used in this paper. We note that although a Markov perfect equilibrium is also a Nash equilibrium of the game, the opposite might not be true always. An \mpe $(\tsigma)$ satisfies sequential rationality such that for $\mathbb{G}_T$, $\forall i\in\cN$, $\forall t \in \cT$, $\forall h^{i}_t \in \cH^i_t$, where $\cH^i_t$ is the space of all possible mean field trajectories till time $t$, and $\forall {\sigma^{i}}\in \cK^\sigma$,
	\begin{align}
		\label{Eqn:seqeq2}
		&\mE^{\tsigma^{i} \tsigma^{-i}}\left[\sum_{n=t}^T \delta^{n-t} R\left(X_n^i,A_n^i,Z_n\right)|z_{1:t},x_{1:t}^i \right]\nn\\&\geq \mE^{{\sigma}^{i} \tsigma^{-i}}\left[\sum_{n=t}^T \delta^{n-t} R\left(X_n^i,A_n^i,Z_n\right)|z_{1:t},x_{1:t}^i \right].   
	\end{align}
	
\subsection{A solution concept: \mfe}
	\mfe can be defined as the combination of the optimal policy $\tsigma\coloneqq\{\tsigma_t\}_{t\in\cT}$ and the mean field states $z\coloneqq\{z_t\}_{t\in\cT}$ that satisfy the following:
	\begin{enumerate}
		\item A policy $\tsigma$ for some $z_{1:T}$ such that
		\begin{align}
			\label{Eqn:seqeq12}
			&\mE^{\tsigma^{i}}\left[\sum_{k=t}^T \delta^{k-t} R\left(X_k^i,A_k^i,Z_k\right)\vert z_{1:T},x_{1:t}^i \right]\nn\\
			&\geq \mE^{\sigma^i}\left[\sum_{k=t}^T \delta^{k-t} R\left(X_k^i,A_k^i,Z_k\right)|z_{1:T},x_{1:t}^i \right].   
		\end{align}
		\item A function $\Phi[z]:\cS_z\to 2^{\cS_\sigma}$ such that
		\begin{align}
			\Phi\cm{z} = \{\tsigma\in S_\sigma\colon\tsigma\text{ is optimal for }z\}
		\end{align}
		\item A mapping $\Lambda\colon\cS_\sigma\to\cS_z$ with $\tsigma\in S_\sigma$, $z=\Lambda\cm{\tsigma}$ is constructed recursively as,
		\begin{align}
			z_{t+1}(y) =\sum_{x_t,a_t} \tau\left(y|x,a,z_t\right)\tsigma_t(a|z_t,x_t)z_t(dx_t)
		\end{align} 
	\end{enumerate}
	Then, the pair $\cm{\tsigma,z}$ can be called a \mfe which is a good approximation for the \mpe when the population grows large.

\subsection{A methodology to compute \mfe }
\label{sec:methodology}
In this section, we summarize the backward recursive methodology based on sequential decomposition that would be used to compute the \mpe. It is worth noting that,~\cite{Vasal2019} provides a sequential decomposition algorithm used for the case when model of the \mdp us known. Here, we modify the algorithm so that it could be used for the model-free case as well. We consider Markovian equilibrium strategies of player $i$ which depend on the common information $z_t$ at time $t$, and on its current private type $x_t^i$. Equivalently, player $i$ takes action of the form $A_t^i\sim \sigma_t^i(\cdot|z_t,x_t^i)$. Similar to the common agent approach~\cite{Nayyar2013}, an alternate and equivalent way of defining the strategies of the players is as follows. We consider a common fictitious agent that views the common information $z_t$ and generates a prescription function $\gamma_t^i:\cX\to\cP(\cA)$ as a function of $z_t$ through an equilibrium generating function $\theta_t^i:\cZ\to(\cX\to\cP(\cA))$ such that $\gamma_t^i = \theta_t^i[z_t]$. Then action $A_t^i$ is generated by applying this prescription function $\gamma_t^i$ on player $i$'s current private information $x_t^i$, i.e. $A_t^i\sim \gamma_t^i(\cdot|x_t^i)$. Thus $A_t^i\sim \sigma_t^i(\cdot|z_{t},x_t^i) = \theta_t^i[z_t](\cdot|x_t^i)$.

We are only interested in symmetric equilibria of such games such that $A_t^i\sim \gamma_t(\cdot|x_t^i) = \theta_t[z_t](\cdot|x_t^i)$ i.e. there is no dependence of $i$ on the strategies of the players.

For a given symmetric prescription function $\gamma_t = \theta[z_t]$, the statistical mean field $z_t$ evolves according to the discrete-time McKean Vlasov equation, $\forall y\in\cX$:
\begin{align}
	\label{eq:z_update}
	z_{t+1}(y) =\sum_{x\in\cX}\sum_{a\in \cA} z_t(x)\gamma_t(a|x)\tau\left(y|x,a,z_t\right),
\end{align}
which implies
\begin{align}
	\label{Eqn:z_update1}
	z_{t+1}= \phi\left(z_t,\gamma_t\right).
\end{align}

\subsection{Backward recursive algorithm for $\mathbb{G}_{T}$}
\label{sec:fhbr}
This section summarizes the proposed a novel model-free algorithm to compute the optimum policy function $\ttheta_t$ as a function of mean field $z_t$ where equilibrium generating function $(\ttheta_t)_{t\in[T]}$ is defined as $\ttheta_t:\cZ\to\{\cX\to\mathcal{P}(\cA) \}$, and for each $z_t $, we generate $\tgamma_t = \ttheta_t[z_t]$. In addition, we generate an action value function $Q_t$ defined as $Q_t:\Pi\times\cZ\times\cX\times\cA\to\mathbb{R}$ that captures the expected sum of returns at time $t$ following certain action from a state and then continuing with the optimal policy $\tsigma_{t+1}$ from time $t+1$ onwards. As per our knowledge, this is the first attempt to solve a fixed point equation using the action value function in a model-free algorithm and determining the corresponding equilibrium policy when the mean field is non-stationary. The algorithm can be summarized as follows:

\begin{enumerate}
	\item Initialize $\forall z_{T+1}, x_{T+1}^i\in \cX, A_{T+1}^i\in\cA, \tgamma_{T+1}\in\Pi$,
	\begin{align}
		V_{T+1}\sq{z_{T+1},x_{T+1}^i} &\defeq 0,\label{Eqn:Intial_V}\\
		\ttheta_{T+1}\sq{z_{T+1}} &\defeq 0\label{Eqn:Intial_theta}.
	\end{align}
   
	\item For $t = T, T-1 \ldots 1$ and  $\forall z_t$, $\ttheta_t\sq{z_t} $ is generated through the following steps. 
	\begin{enumerate}
		\item Compute $Q_t$, $\forall x_t^i\in\cX$, $\forall a_t^i\in\cA$, and $\forall \tgamma_t\in\Pi$ as,
		\begin{align}
			Q_t(z_t,x_t^i,&a_t,\tgamma_t) \defeq R\cm{z_t,x_t^i,a_t^i} 	+ \nn\\
			&\delta \mE\big[V_{t+1}\cm{z_{t+1}, X_{t+1}^i} | z_t,x_t^i\big],\label{Eqn:Qdef}
		\end{align}
		where the expectation in~\eqref{Eqn:Qdef} is with respect to the random variable $X_{t+1}^i$ through the measure $\tau\left(x_{t+1}^i\vert x_t^i,a_t^i,z_t\right)$. The mean state $z_{t+1}=\phi\left(z_t,\tgamma_t\right)$.

		\item Set $\ttheta_t[z_t] = \tgamma_t$, where $\tgamma_t$ is the solution to the following fixed-point equation at all $x_t^i\in \cX$ and $\forall i \in \cN$
		\begin{align}
			\label{Eqn:fixed_point}
			 &\tgamma_t\cm{\cdot\vert x_t^i}\in\nn\\
			 &\arg\max_{\gamma_t\cm{\cdot\vert x_t^i}}\mE^{\gamma_t\cm{\cdot\vert x_t^i}}\sq{Q_t\cm{z_t,x_t^i,A_t^i,\tgamma_t} \vert z_t,x_t^i}
		\end{align}
		where expectation in~\eqref{Eqn:fixed_point} is with respect to random variable $A_t^i$ through the measure $\gamma_t(a_t^i|x_t^i)$. 
		
		\item The value function $V_{t}$ is computed $\forall x_t^i\in\cX$ as,
		\begin{align}
			\label{Eqn:Vdef}
			V_t\cm{z_t,x_t^i}=\mE^{\tgamma_t\cm{\cdot\vert x_t^i}}\sq{Q_t\cm{z_t,x_t^i,A_t^i,\tgamma_t} \vert z_t,x_t^i}
		\end{align}
	\end{enumerate}
\end{enumerate}
Then, an equilibrium strategy is defined as 
\begin{align}
\label{Eqn:sigma_fh}
	\tilde{\sigma}_t^i\left(a_t^i|z_{1:t},x_{1:t}^i\right) = \ttheta[z_t]\left(a_t^i|x_t^i\right),
\end{align}
where $ \ttheta[z_t] = \tilde{\gamma}_t$. The proof for the existence of the solution to the fixed point equation in \eqref{Eqn:fixed_point} which is the \mpe of the game has already been provided in~\cite{Vasal2019} and will be revisited when we prove the convergence of our algorithm.

%% file: 4Algorithm.tex
In this section, we describe our proposed \rl algorithm that computes the optimal policy which maximizes the expected sum of returns as specified in~\eqref{Eqn:j_func}. The optimal policy $\tsigma$ is defined for a discretized set of mean states $z\in\cZ$, given as $\ttheta\sq{z}$. At any time $t$, and for any current mean state $z_t\in\cZ$, $\ttheta_t\sq{z_t}$ maps to a function $\tgamma_t$ that prescribes the probabilistic action $a_t^i$, an agent $i$ should take, given the state $x_t^i$. 
	
We implement the \rl algorithm based on Expected Sarsa~\cite{VanSeijen} and without the explicit knowledge of the the transition probabilities $\tau(x_{t+1}^{i}\vert x_t^{i},a_t^i)$. The algorithm basically computes the $Q_t$-values at each instant and then learns the optimal policy $\tgamma_t$ by solving the fixed point equation in~\eqref{Eqn:fixed_point}. We use Expected Sarsa to update the $Q_t$ values, from the current reward $r_t$ and $V_{t+1}$, the value at the future state. This update can be expressed as 
\begin{align}
	\label{Eqn:Sarsa_update}
	Q_t\cm{z_t,x_t,a_t,\tgamma_t}=&\cm{1-\alpha}Q_t\cm{z_t,x^i_t,a_t^i} +\nn\\
	&\alpha\cm{r_t+\delta V_{t+1}\cm{z_{t+1},x^i_{t+1}}}
	\end{align}
The $Q$-values are a function not only of the states and the actions but also of the current mean field state $z_t$ and the current optimal policy $\ttheta\sq{z_t}=\tgamma_t$. The current optimal policy and current mean state determine the next mean state $z_{t+1}$ which determines the value function $V_{t+1}\cm{z_{t+1},x^i_{t+1}}$ at the future state. Therefore, the functions $Q_t$ is defined over all possible equilibrium policies $\tgamma_t\left(\cdot\vert x_t\right)\in \Pi$, where $\Pi$ is the space of all possible strategies from a given state $x_t$. The value function $V_{t+1}\cm{z_{t+1},x^i_{t+1}}$ is determined using functional approximation. In addition, due to the non-stationarity of the mean states, the equation in~\eqref{Eqn:fixed_point}, which is used to solve for the optimal policy, is not just a single step optimization, but a fixed point equation which needs to be solved with repeated iterations. In our paper, we use a policy gradient approach to solve for the optimal policy at each mean state and for every time iteration repeatedly. The entire \rl algorithm described here is summarized in Algorithm~\ref{alg:Evaluation}. 

At each time instant $t$, the policy iteration algorithm computes the equilibrium policy based on the action value function $Q_t$ through a policy gradient approach. In other words, the solution to the fixed point equation in \eqref{Eqn:fixed_point} is the policy where $Q_t$ has the highest gradient. Given that the the function $Q_t$ is a function of the optimal policy itself, the new found policy changes the $Q_t$. Therefore, this process is repeated over several iterations in order to arrive at the required prescription function $\tgamma_t$. This is repeated at all the mean states $z_t\in\cZ$ so that we get the final equilibrium function $\ttheta\left[z\right]$.

\begin{algorithm}[t]
	\label{alg:Evaluation}
	\DontPrintSemicolon
	\SetAlgoLined
	\KwIn
	{\\
	L: Batch Size for Sarsa\\
	I: Policy Iterations\\
	T: Time Length
	}
	\vspace{.1cm}
	\KwOut
	{\\
		$\tsigma$: Optimal Policy
	} 
	Initialize: $V_{T+1}$\;
	Initialize: $\ttheta_{T+1}$\;
	\For{t = T\ldots 1}
	{	
		\For{$z_t\in\cZ$, $\tgamma_t\in\Pi$}
		{
			Next mean state: $z_{t+1}\sim\phi\cm{z_t,\tgamma_t}$\;
			Optimum policy: $\tgamma_{t+1} = \ttheta_{t+1}\sq{z_{t+1}}$\;
			\For{$l = 1,2,\ldots L$}
			{	
				\For{$\left(x_l^i,a_l^i\right)\in\cX\times\cA$}
				{
					Sample: $x_{l+1}^i\sim \tau\left(.\vert x_l^i, a_l^i, z_t\right)$\;
					Sarsa Target: $G = R(x_l^i,a_l^i,z_t) +\delta V_{t+1}\left(z_{t+1},x_{l+1}^i\right)$\;
					$Q_t\left(z_t,x_l^i, a_l^i,\tgamma_t\right) = \left(1-\alpha\right)Q_t\left(z_t,x_l^i, a_l^i,\tgamma_t\right) + \alpha G$\;  
				}
			}
		}
		\For{$z\in\cZ$}
		{
			\For{$n=1\ldots I$}
			{
				$\tgamma_n$ = PG $\left(Q_t\left(z_t,\cdot,\cdot,\tgamma_{n-1}\right)\right)$\;
				Increment: $n = n+1$\;
			}
			$\ttheta\sq{z_t}=\tgamma_n$
		}
		\For{$z\in\cZ$}
		{
			\For{$x^i\in\cX$}
			{	
				$\tgamma = \ttheta\sq{z_t}$\;
				$V_t\cm{z_t,x^i} = \mE^{\tgamma}\sq{Q_t\cm{z_t,x^i,A^i,\tgamma}}$\;
			}
		}
	}	
	$\tsigma = \ttheta[z_t] \ \forall z_t$\;
	\KwResult{$\tsigma$}
	\caption{Equilibrium Policy}
\end{algorithm}

%% file: 5Convergence.tex
\label{sec:Convergence}

In this section, we prove the convergence of the proposed \rl algorithm to the equilibrium strategy of the statistical \mfg. Using backward recursion and sequential decomposition, the \rl algorithm is able to arrive at the equilibrium strategy for player $i$ at each time $t$. In other words, we show that the a player $i$ has no incentive to deviate from the equilibrium strategy $\tsigma$ given that the other players are playing the equilibrium strategy. Before proving convergence, we establish two lemmas related to the main theorems. In the first lemma we show that the value function $V_t$ captures the expected sum of rewards accumulated by playing the $\tsigma$ at time $t$ by the $i^{th}$ player. We follow it with the proof of the next lemma establishing the optimality of the $V$ value over the expected sum of rewards accumulated by playing any other strategy other than $\tsigma$. 

\begin{lemma} 
	\label{Lemma:V_optimal_returns}
	$\forall t\in\sq{T}$, $\forall z_t$, $x_t^i\in\cX$,
	\begin{align}
		V_t\cm{z_t,x_t^i} &= \mE^{\tsigma_{t:T}}\left[\sum_{k=t}^{T}\delta^{k-t}R\left(x_k^i,A_k^i,z_k\right)\vert z_t,x_t^i\right]\\
		&=J_t^{\tsigma}
	\end{align}
	where $\tsigma_t$ is the equilibrium policy at time $t$ and $J_t^{\tsigma}$ is the accumulated optimal returns from $t$ till $T$ by following the equilibrium policy. 

	\begin{proof}
		We prove the lemma using the theory of mathematical induction. 

		At $t=T$, from~\eqref{Eqn:Vdef},
		\begin{align}
			\label{Eqn:lemma1_eqn1}
			V_T\cm{z_T, x_T^i} =& \mE^{\tsigma_T}\sq{Q_T\cm{z_T, x_T^i, A_T^i, \tsigma_T\cm{\cdot\vert \cdot,z_t}}}\\
			=& \mE^{\tsigma_T}\sq{R\cm{x_T^i, A_T^i,z_T}},
		\end{align}
		which is true from~\eqref{Eqn:Qdef}. For any mean state $z_t$, $\tgamma_t=\tsigma_t\cm{\cdot\vert \cdot,z_t}$, therefore, we have,
		\begin{align}
			\label{Eqn:lemma_eqn2}
			V_T\cm{z_T,x_T^i} = \mE^{\tgamma_T}\sq{R\cm{x_T^i,A_T^i}}.
		\end{align}
		which is the maximum returns the agents can receive at $t=T$ because $\tgamma_t$ is the solution to the fixed point equation in~\eqref{Eqn:fixed_point} that maximizes~\eqref{Eqn:lemma_eqn2}.

		Now assuming that the proposition is true for $t = t+1$, we get,
		\begin{align} 	 
			&V_{t+1}\cm{z_{t+1},x_{t+1}^i} =J_{t+1}^{\tsigma}\\
			&=\mE^{\tsigma_{t+1:T}}\left[\sum_{k=t+1}^{T}\delta^{k-(t+1)}R\left(x_k^i,A_k^i,z_k\right)\vert z_{t+1},x_{t+1}^i\right]\label{Eqn:Lemma1_true_for_t+1}\\
		\end{align}
		At time $t = t$, we have,
		\begin{subequations}
			\begin{align}
				V_t\cm{z_t, x_t^i} = &\mE^{\tsigma_t}\sq{Q_t\cm{z_t, x_t^i, A_t^i, \tsigma_t\cm{\cdot\vert \cdot,z_t}}\vert z_t,x_t^i}\label{Eqn:Lemma1_eqn1}\\
								 = &\mE^{\tsigma_t}[R\cm{x_t^i, z_t,A_t^i}+\nn\\
								 &\delta V_{t+1}\cm{z_{t+1},x_{t+1}^i}\vert z_t,x_t^i]\label{Eqn:Lemma1_eqn2}
			\end{align}
		\end{subequations}
		\eqref{Eqn:Lemma1_eqn1} is from the definition in~\eqref{Eqn:Vdef} while~\eqref{Eqn:Lemma1_eqn2} is from the definition in~\eqref{Eqn:Qdef}. Using the assumption in~\eqref{Eqn:Lemma1_true_for_t+1}, we get,
		\begin{align}
			&V_t\cm{z_t, x_t^i} = \mE^{\tsigma_t}\Bigg[R\cm{x_t^i, A_t^i, z_t}+\nn\\
			&\delta  \mE^{\tsigma_{t+1:T}}\left[\sum_{k=t+1}^{T}\delta^{k-(t+1)}R\left(x_k^i,z_k\right)\vert z_{t+1},x_{t+1}^i\right]\nn\\
			&\quad\quad\quad\quad\quad\quad\quad\quad\quad\quad\quad\quad\vert z_t,x_t^i\Bigg]
		\end{align}
		Using the the expression for expected sum of returns in~\eqref{Eqn:j_func}, we get,
		\begin{align}
            V_t\cm{z_t, x_t^i}    = \mE^{\tsigma_t}[R\cm{x_t^i, z_t, A_t^i}+\delta  J_{t+1}]
		\end{align}

	\end{proof}
\end{lemma}

\begin{lemma} \label{Lemma:V_is_better}
	$\forall i\in\cN$,$t\in\sq{T}$,$\forall z_t,x_t^i\in\cX,a^i_t\in\cA$,
	\begin{align}
		\label{Eqn:lemma2_to_prove}
		V_t\left(z_t,x_t^i\right) \geq \mE^{\sigma_{t}^i,\tsigma_{t}^{-i}}\left[Q_{t}\left(z_{t}, x_{t}^i, A_{t}^i,\tgamma_{t}\right) \vert z_t,x_t^i\right]
	\end{align}
	where $z_{t+1} = \phi\left(z_t,\tsigma_t\left(\cdot\vert z_t,\cdot\right)\right)$.

	\begin{proof}
		Given that at time $t+1$ the equilibrium policy is $\tsigma_{t+1}$ is the solution to the fixed point equation in~\eqref{Eqn:fixed_point}, let us assume that the player $i$ plays a different policy $\hat{\sigma}_{t+1}$ such that $\forall x^i_{t+1}\in\cX$,
		\begin{align}
			\label{Eqn:lemma2_not_in}
			\hat{\gamma}_{t}\cm{\cdot\vert x_{t+1}^i}\notin\arg\max_{\gamma_t} \mE\sq{Q_{t}\cm{z_{t},x_{t}^i,A_{t}^i,\tgamma_{t}} | z_{t},x_{t}^i}
		\end{align}
		with $\hat{\sigma}_{t}\cm{\cdot\vert z_{t},x_{t}^i}=\hat{\gamma}_{t}\cm{\cdot\vert x_{t}^i}$.

		Now, the equation on the other side of the inequality in\eqref{Eqn:lemma2_to_prove} can be changed assuming $\hat{\sigma}$ as the suboptimal policy as follows:
		\begin{subequations}
			\begin{align}
				&\mE^{\sigma_{t}^i,\tsigma_{t}^{-i}}\sq{Q_{t}\cm{z_{t+1}, X_{t}^i,A_{t}^i,\tgamma_{t}}\big\vert z_t,x_t^i,a_t^i}\nn\\
				=&\mE^{\hat{\gamma}_{t}^i,\tsigma_{t}^{-i}}\sq{Q_{t}\cm{z_{t}, x_{t}^i,A_{t}^i,\tgamma_{t}}\big\vert z_t,x_t^i,a_t^i}
			\end{align}
		\end{subequations}
		Considering~\eqref{Eqn:lemma2_not_in}, we can proceed as,
		\begin{subequations}
			\begin{align}
				&\mE^{\hat{\gamma}_{t}^i,\tsigma_{t}^{-i}}\sq{Q_{t}\cm{z_{t}, X_{t}^i,A_{t}^i,\tgamma_{t}}\big\vert z_t,x_t^i,a_t^i}\nn\\
				\leq&\mE^{\tgamma_{t}^i,\tsigma_{t}^{-i}}\sq{Q_{t}\cm{z_{t+1}, X_{t}^i,A_{t+1}^i,\tgamma_{t}}\big\vert z_t,x_t^i,a_t^i}\\
				=&V_t\left(z_t,x_t^i\right)
			\end{align}
		\end{subequations}
		where the last statement is from the definition in~\eqref{Eqn:Qdef}.
	
	\end{proof}
\end{lemma}

\begin{theorem}\label{Thm:main_theorem}
	A strategy $(\tilde{\sigma})$ constructed from the above algorithm is an \mfe of the game. i.e
	\begin{align}
		 &\mE^{\tsigma}\sq{\sum_{k = t}^T\delta^{k-t}R\cm{X_k^i,A_k^i,Z_k}\big\vert z_{1:t},x^i_{1:t}}\nn\\
		 &\geq\mE^{\sigma^i,\tsigma^{-i}}\sq{\sum_{k = t}^T\delta^{k-t}R\cm{X_k^i,A_k^i,Z_k}\big\vert z_{1:t},x^i_{1:t}}\label{Eqn:main_theorem}	
	\end{align}
	\begin{proof}
		We prove it through the technique of mathematical induction and will use the results that were proved before in Lemma~\ref{Lemma:V_optimal_returns} and Lemma~\ref{Lemma:V_is_better}.
		
		For the base case, we consider $t = T$. The expected sum of returns, when the player $i$ follows the equilibrium policy $\tsigma$ is given as
		\begin{align}
			J_T &= \mE^{\tsigma_T}\sq{R\cm{x_T^i,A_T^i,z_T}\big\vert z_{1:T},x^i_{1:T}}\\
				& = V_T\cm{z_T,x_T^i},
		\end{align}
		which is true from Lemma~\ref{Lemma:V_optimal_returns}.

		Now, from Lemma~\ref{Lemma:V_is_better}, we have,
			\begin{align}
				V_T\geq& \mE^{\sigma_{T}^i,\tsigma_{T}^{-i}}\left[Q_{T}\left(z_{T}, x_{T}^i, A_{T}^i,\tgamma_{T}\right) \vert z_t,x_t^i\right]\\
				=&\mE^{\sigma_T^i,\tsigma_T^{-i}}\sq{R\cm{x_T^i,A_T^i,z_T}\big\vert z_{1:T},x^i_{1:T}}
			\end{align}
		Assuming that the condition in~\eqref{Eqn:main_theorem} holds at $t = t+1$, we get,
		\begin{align}
			\begin{split}
				&\mE^{\tsigma}\sq{\sum_{k = t+1}^T\delta^{k-t-1}R\cm{x_k^i,A_k^i,z_k}\big\vert z_{1:t+1},x^i_{1:t+1}}\\
				&\geq\mE^{\sigma^i\tsigma^{-i}}\sq{\sum_{k = t+1}^T\delta^{k-t- 1}R\cm{x_k^i,A_k^i,z_k}\big\vert z_{1:t+1},x^i_{1:t+1}}\label{Eqn:thm1_assumption}
			\end{split}
		\end{align}
		We need to prove that the expression in~\eqref{Eqn:main_theorem} holds for $t=t$ as well. Let us represent the left hand side of~\eqref{Eqn:main_theorem} as $L$, i.e.
		\begin{align}
			\label{Eqn:thm1_t}
			L = \mE^{\tsigma}\sq{\sum_{n = t}^T\delta^{k-t}R\cm{X_k^i,A_k^i,Z_k}\big\vert z_{1:t},x^i_{1:t}}
		\end{align}
		The expectation at $t+1$ is independent of the rewards at time $t$, therefore, we can rewrite~\eqref{Eqn:thm1_t} as
		\begin{align}
			L &=\mE^{\tsigma_t}\Bigg[R\cm{x_t^i,a_{t}^i,z_t} + \nn\\
			&\delta\mE^{\tsigma_{t+1:T}}\Bigg[\sum_{k = t+1}^T\delta^{k-t-1} R\cm{X_k^i,A_k^i,Z_k}\big\vert z_{1:t+1},x^i_{1:t+1}\Bigg]\Bigg]\label{Eqn:thm1_a}
		\end{align}
		Using Lemma~\ref{Lemma:V_optimal_returns}, and the definition of $V_t$ in~\eqref{Eqn:Vdef} we get,
		\begin{subequations}
			\begin{align}
				L=&\mE^{\tsigma_t}\sq{R\cm{x_t^i,a_{t}^i,z_t} + \delta V_{t+1}\cm{z_{1:t+1},x^i_{1:t+1}}}\\
				=&\mE^{\tsigma_t}\sq{Q_t\cm{z_t,x_t^i,a_t^i,\tsigma_t\cm{\cdot\vert z_t,\cdot}}}\\
				=&V_t\cm{z_t,x_t^i}
			\end{align}			
		\end{subequations}
		Now, from Lemma~\ref{Lemma:V_is_better},
		\begin{subequations}
			\begin{align}
				L\geq&\mE^{\sigma_{t}^i,\tsigma_{t}^{-i}}\sq{Q_{t}\cm{z_{t}, x_{t}^i, A_{t}^i,\tgamma_{t}} \vert z_t,x_t^i}\\
				=&\mE^{\sigma_{t}^i,\tsigma_{t}^{-i}}\sq{R(x_t^i,A_t^i,z_t) + \delta V_{t+1}\cm{z_{t+1},X_{t+1}^i}\vert z_t,x_t^i}\label{Eqn:Thm1_eqn6}
			\end{align}
		\end{subequations}
		which is true as per the definition in~\eqref{Eqn:Qdef}. The expression can now be expanded using Lemma~\ref{Lemma:V_optimal_returns} and use the assumption we made in~\eqref{Eqn:thm1_assumption}.
		\begin{subequations}
			\begin{align}
				L\geq&\mE^{\sigma_{t}^i,\tsigma_{t}^{-i}}\Bigg[R(z_t,x_t^i,a_t^i) + \nn\\
				&\delta \mE^{\tsigma}\sq{\sum_{k=t}^{T}\delta^{k-t}R\left(x_k^i,A_k^i,z_k\right)\vert z_t,x_t^i}\Bigg]\\
				\geq&\mE^{\sigma_{t}^i,\tsigma_{t}^{-i}}\Bigg[R(z_t,x_t^i,a_t^i) + \nn\\
				&\delta\mE^{\sigma^i\tsigma^{-i}}\sq{\sum_{k = t+1}^T\delta^{k-t- 1}R\cm{x_k^i,A_k^i,z_k}\big\vert z_{1:t+1},x^i_{1:t+1}}\Bigg]\\
				=&\mE^{\sigma^i,\tsigma^{-i}}\sq{\sum_{k = t}^T\delta^{k-t}R\cm{X_k^i,A_k^i,z_k}\big\vert z_{1:t},x^i_{1:t},a_{1:t}^i}\label{Eqn:Thm1_eqn10}
			\end{align}
		\end{subequations}
	\end{proof}
\end{theorem}	

\begin{theorem}
	Let $\tilde{\sigma}$ be an \mfe of the mean field game. Then there exists an equilibrium generating function $\ttheta$ that satisfies \eqref{Eqn:fixed_point} in the backward recursion algorithm such that $\tilde{\sigma}$ is defined using $\ttheta$.

	\begin{proof}
		Given that $\tsigma$ is the \mfe of the game, the equilibrium generating function such that
		\begin{align}
			\ttheta[z] = \tsigma\cm{\cdot\vert z,\cdot} \quad \forall z
		\end{align}
		At time $t$, let us assume that $\tgamma_t\cm{=\ttheta\sq{z_t}}$ be the prescription function that gives the equilibrium action but does not satisfy the fixed point equation in~\eqref{Eqn:fixed_point}, i.e.
		\begin{align}
			\tgamma_t\notin\arg\max_{\gamma_t}\mE^{\gamma_t}\sq{Q_t\cm{z_t,x_t^i, A_t^i,\tgamma_t}}
		\end{align}  
		Let us assume that $\hat{\gamma}_t$ be a solution to the fixed point equation in the \rl algorithm, then
		\begin{align}
			\mE^{\hat{\gamma}_t}\sq{Q_t\cm{z_t,x_t^i,A_t^i,\tgamma_t}}&\geq\mE^{\tgamma_t}\sq{Q_t\cm{z_t,x_t^i,A_t^i,\tgamma_t}}
		\end{align}
		But, we know
		\begin{subequations}
			\begin{align}
				&\mE^{\hat{\gamma}_t}\sq{Q_t\cm{z_t,x_t^i,A_t^i,\tgamma_t}}\\
				&=\mE^{\hat{\sigma}_t}\sq{R_t\cm{x_t^i,A_t^i,z_t}+V_{t+1}\cm{z_{t+1},x^i_{t+1}}}\\
				&=\mE^{\hat{\sigma_t},\tsigma_{t+1:T}}\sq{\sum_{n = t}^T\delta^{n-t}R\cm{X_n^i,A_n^i,Z_n}\big\vert z_{1:t},x^i_{1:t},a_{1:t}^i}\\
				&\geq\mE^{\tgamma_t}\sq{Q_t\cm{z_t,x_t^i,A_t^i,\tgamma_t}}\\
				&=V_t\cm{z_t,x_t^i}
			\end{align}
		\end{subequations}
		which is contradictory on the consideration that $\tsigma$ is a \mpe of the game.
	\end{proof}
\end{theorem}	

\begin{assumption}
	\label{Assume:existence}
	Let the reward function $R\left(x_t^i, a_t^i, z_t\right)$ and the state transition matrix $\tau\left(x_{t+1}^i\vert x_t^i,a_t^i,z_t\right)$ be continuous functions in $z_t$.
\end{assumption}	
\begin{theorem}
	Under the assumption~\ref{Assume:existence}, there exists a solution to the fixed point equation in~\eqref{Eqn:fixed_point}.
	
	\begin{proof}
		It has already been shown that when the reward function is bounded, there exists a \mfe for finite horizon games. It can also been shown that the solutions to the fixed point equations are \mfe of the game. Thus there exists an equilibrium policy for the game achieved at each time $t$.
	\end{proof}
\end{theorem}

%% file: 6Numerical_problems.tex
\subsection{Security of cyber-physical system: Malware Spread}
\label{sec:example}

We consider a security problem in a cyber physical network with positive externalities. It is a discrete version of the malware problem presented in~\cite{huang2017mean,huang2016mean, huang2017mean1, jiang2010bad}. Some other applications of this model include flu vaccination, entry and exit of firms, investment, network effects. In this model, we suppose there are large number of cyber-physical nodes where each node has a private state $x_t^i\in\{0,1\}$ where $x_t^i= 0$ represent `healthy' state and $x^i_t= 1$ is the infected state. Each node can take an action $a_t^i\in\{0,1\}$, where $a_t^i= 0$ implies ``do nothing" whereas $a_t^i=1$ implies ``repair". The dynamics$\left(=Q_x\left(\cdot\vert\cdot\right)\right)$ are given by

\[   
    x_{t+1}^i = 
     \begin{cases}
        x_t^i+ (1-x_t^i)w_t^i \quad\text{for}\quad a_t^i = 0\\
       \quad\quad 0 \ \quad\quad\quad\quad\quad\text{for}\quad a_t^i = 1 
     \end{cases}
\]

where $w_t^i \in \{0,1\}$ is a binary valued random variable with $P(w_t^i = 1) = q$ representing the probability of a node getting infected. Thus if a node doesn't do anything, it could get infected with certain probability, however, if it takes repair action, it comes back to the healthy state. 
Each node gets a reward 
\begin{align}
	r(x^i_t,a^i_t,z_t) =-(k+z_t(1))x^i_t-\lambda a_t^i.
\end{align}
where $z_t(1)$ is the mean field population state being 1 at time $t$, $\lambda$ is the cost of repair and $(k+z_t(1))$ represents the risk of being infected.
We pose it as an infinite horizon discounted dynamic game.
We consider parameters $k= 0.2,\lambda= 0.5, \delta = 0.9, q=0.9$ for numerical results presented in Figures~1-3.

\begin{figure}[htbp] 
   \centering
   \includegraphics[width=.35\textwidth]{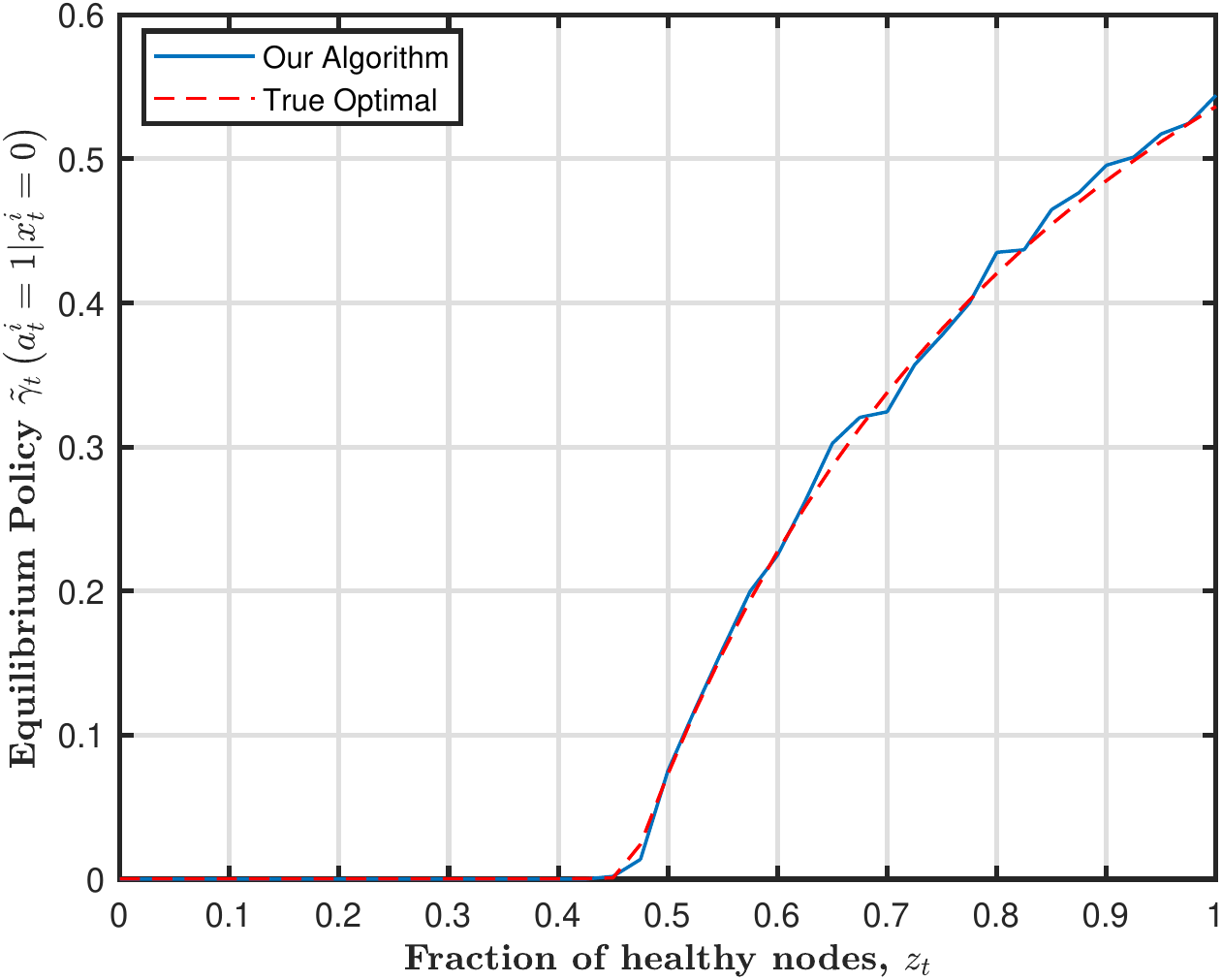} 
   \caption{$\gamma(1|0)$: Probability of choosing action 1, given $x^i = 0 $}
   \label{fig:Figure_P_01}
\end{figure}
\begin{figure}[htbp] 
	\centering
	\includegraphics[width=.35\textwidth]{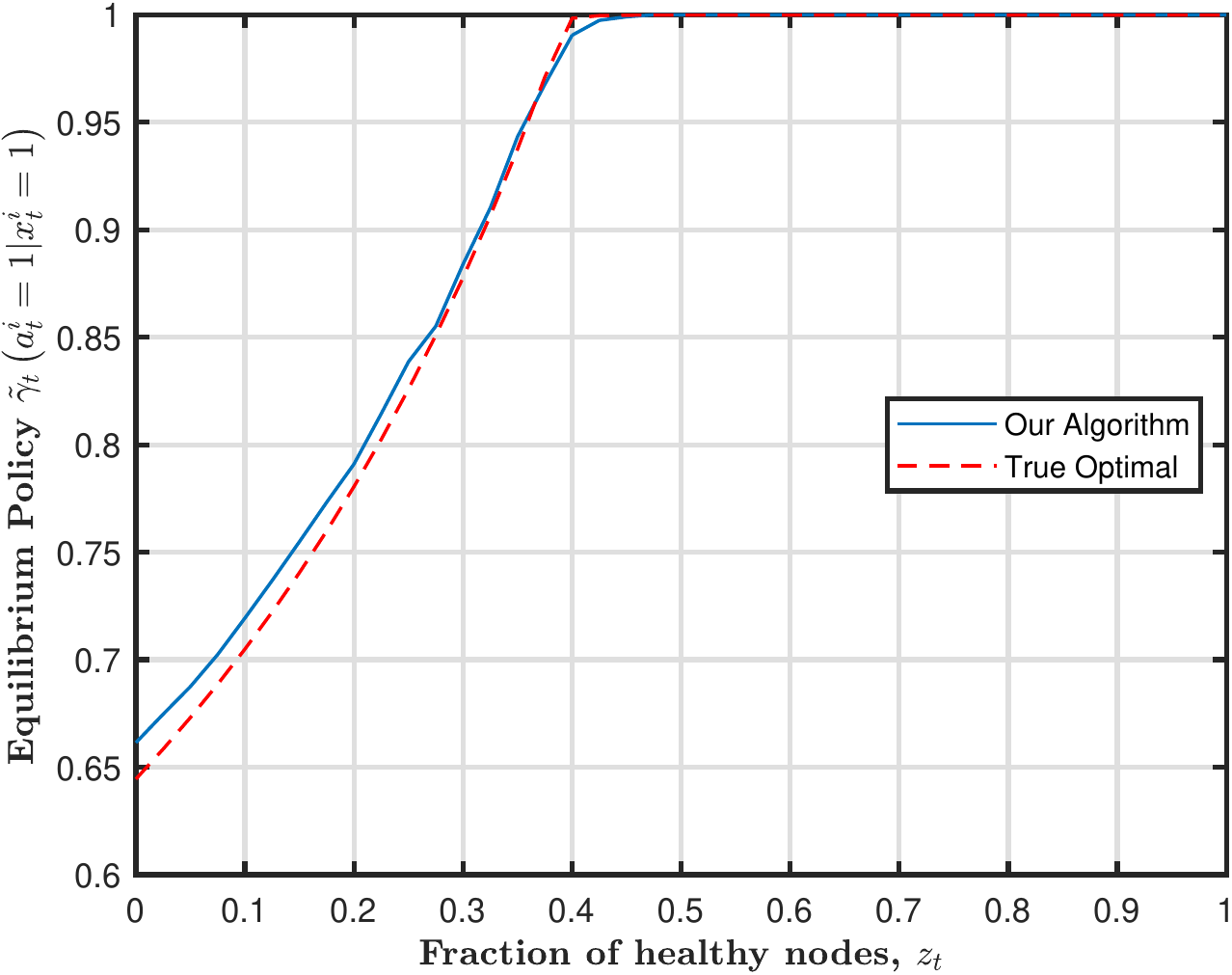} 
	\caption{$\gamma(1|1)$: Probability of choosing action 1, given $x^i = 1 $}
	\label{fig:Figure_P_11}
\end{figure}

 The learning parameter $\alpha$ for the sarsa update was set at $0.1$. The overall length of the time-horizon $T$ was chosen to be $60$ iterations long.

 \figref{fig:Figure_P_01} and \figref{fig:Figure_P_11} show the equilibrium policies $\tgamma$ at different values of the mean state $z_t$ for states $x_t^i=0$ and $x_t^i=1$. The plotted graphs are the probabilities with which we choose action $a_t^i =1$. The plots of our algorithm are compared across the true strategy that was obtained by assuming the knowledge of the dynamics of \mdp and then solving the fixed point equation. The strategies estimated using the proposed \rl algorithm coincides with the true strategies establishing the accuracy of our algorithm.

%% file: 7Conclusion.tex
We considered a finite horizon discrete-time sequential \mfg with infinite homogeneous players. The players had access to their private type and the common information of mean population state. A fixed point decomposition method was suggested in an earlier paper that computes the equilibrium strategy at different mean states but with the knowledge of the dynamics of the \mdp. Here, we proposed a \rl algorithm that employs Expected Sarsa to learn the dynamics of the game and solve the fixed point equation, iteratively, to arrive at the equilibrium strategy. In the end, we implement our algorithm on a practical cyber-physical application to demonstrate that the algorithm does converge to the same optimal policy that was obtained when dynamics of the game was known. We also analytically show the convergence of our algorithm to the \mfe of the game. To the best of our knowledge, this is the first RL algorithm to learn optimal policies of non-stationary mean field games.